\documentclass[intlimits,twoside,a4paper]{article}

\usepackage{amsmath,amssymb}
\usepackage{graphicx}
\usepackage{wrapfig}

\usepackage[T2A]{fontenc}
\usepackage[cp1251]{inputenc}
\usepackage[eqsecnum]{cmpj2}
\issue{2012}{15}{2}{23803}

\doinumber{10.5488/CMP.15.23803}

\title[MC simulation of a slit-system]%
{Monte Carlo simulation of the electrical properties of electrolytes adsorbed in charged slit-systems%
\thanks{We dedicate this paper to Orest Pizio honouring his valuable contribution to the field of statistical mechanics of fluids.}
}
\author[R. Kov\'acs, M. Valisk\'o, D. Boda]{R. Kov\'acs, M. Valisk\'o, D. Boda\footnote{E-mail: boda@almos.vein.hu}
}
\address{Department of Physical Chemistry, University of Pannonia, P. O. Box 158, Veszpr\'em, Hungary}

\date{Received January 16, 2012, in final form February 21, 2012}

\authorcopyright{R. Kov\'acs, M. Valisk\'o, D. Boda, 2012}

\begin{document}

\maketitle

\begin{abstract}
We study the adsorption of primitive model electrolytes into a layered slit system using grand canonical Monte Carlo simulations.
The slit system contains a series of charged membranes.
The ions are forbidden from the membranes, while they are allowed to be adsorbed into the slits between the membranes.
We focus on the electrical properties of the slit system.
We show concentration, charge, electric field, and electrical potential profiles.
We show that the potential difference between the slit system and the bulk phase is mainly due to the double layers formed at the boundaries of the slit system, but polarization of external slits also contributes to the potential drop.
We demonstrate that the electrical work necessary to bring an ion into the slit system can be studied only if we simulate the slit together with the bulk phases in one single simulation cell.
\pacs 82.45.Mp, 82.45.Gj, 87.10.Rt, 61.20.Qg
\end{abstract}

\section{Introduction}
\label{sec:intro}

Electrical double layers (DLs) formed in an electrolyte near a charged surface have been studied using both simulation \cite{torrie-cpl-65-343-1979,torrie-jcp-1980,torrie-jcp-1982,valleau-jcp-1982,torrie-jpc-1982,valleau-jcp-1984,torrie-jcp-1984,vanmegen-jcp-1980,snook-jcp-1981,jonsson_jpc_1980,boda-jcp-116-7170-2002,henderson-pccp-11-3822-2009} and theoretical \cite{blum_jpc_1977,rosenfeld-prl-63-980-1989,kierlik-pra-42-3382-1990,kierlik-pra-44-5025-1991,yteran_jcp_1990,rosenfeld-pre-55-4245-1997,gillespie02,gillespie03,di-mp-101-2545-2003,di-ea-48-2967-2003,outhwaite-jcsft-1983} methods.
The basis of these studies is the Primitive Model (PM) of electrolytes that represents the solvent as a dielectric continuum.
Simulations necessarily use a finite simulation cell for these studies, where the electrolyte is confined between two charged walls.

If these walls are far enough from each other, the DLs formed at the walls are independent of each other and a charge neutral bulk region is formed in the middle of the simulation cell.
The reference point for the electrical potential then can be set in this bulk region.
If the walls, however, are close to each other so that the two DLs overlap (slit), the bulk region disappears \cite{pizio04,patryk04,pizio05c,yu06,borowko05,pizio05a,buyukdagli_jsm_2011,ibarraarmenta_pccp_2011,martinm_jpcb_113_2009,martinmolina_sm_2010,wang_jcp_2011}.
Slits are generally simulated in the grand canonical (GC) ensemble, where the electrolyte in the slit is in equilibrium with a virtual bulk phase represented by its temperature and the chemical potentials of the ionic species.
The ground of the electrical potential cannot be set in this bulk phase, because it is not present in the simulation cell and Poisson's equation cannot be integrated over it.

The studies for electrolytes confined in a slit considered only a single slit (we will call it the ``lonely slit'') in equilibrium with a bulk in the GC ensemble \cite{pizio04,patryk04,pizio05c,yu06,borowko05,pizio05a,buyukdagli_jsm_2011,ibarraarmenta_pccp_2011,martinm_jpcb_113_2009,martinmolina_sm_2010,wang_jcp_2011}.
In this work we are primarily concerned with the electrical properties of the slit.
Most importantly, we want to calculate the potential difference between a slit and the bulk phase. Namely, we are concerned with the electrical work needed to bring an ion from bulk into the slit.
Therefore, we simulate the slits confined between membranes with bulk electrolytes on the other sides of the membranes.

Furthermore, we consider more than one slit (we will call these ``slit systems'').
Slit systems are first-order models for layered silicate minerals \cite{deville-langmuir-1998,jonsson-2004,jonsson-2005,pegado_jpcm_2008}, porous electrodes \cite{kiyohara07,kiyohara_jcp_2010,kiyohara_jcp_2011,KiyoharaJPCC,kiyohara_smc_2011}, and lyotropic lamellar liquid crystals \cite{jonsson_jpc_1980,ekwall}.
The structure, swelling, and adsorption properties of such materials (e.g. kaolinite, montmorillonite) are subject of extensive experimental and simulation studies \cite{shroll_jcp_1999,chavez-paez,rutkai_cpl_2008,rutkai_jcis_2009,mako_jcis_2010}.
In this work, we focus on the electrical properties of charged slit systems and show concentration, charge, electric field, and potential profiles for different geometrical parameters (width of the slit, width of the membrane), membrane charge, and electrolytes (concentration, ionic charge).

\section{Model of the slit system}
\label{sec:model}

The electrolyte is modeled with the Restricted Primitive Model (RPM).
In this model, the solvent is represented by its dielectric response characterized by the dielectric constant $\epsilon$, while the ions are represented by charged hard spheres interacting through the screened Coulomb + hard sphere pair potential:
\begin{equation}
u(r_{ij}) =
\left\lbrace
\begin{array}{ll}
\infty & \quad \mathrm{for} \quad r_{ij}<R_{i}+R_{j}\,, \\
 \dfrac{1}{4\pi\epsilon_{0}\epsilon} \dfrac{q_{i}q_{j}}{r_{ij}} & \quad \mathrm{for} \quad r_{ij} \geqslant  R_{i}+R_{j}\,,\\
\end{array}
\right.
\label{eq:uij}
\end{equation}
where $r_{ij}$ is the distance of two ions, $q_i$ is the ionic charge ($ q_{i}=z_{i}e$, $z_{i}$ being the valence and $e$ the elementary charge), and $R_{i}$ is the ionic radius.
In the RPM, $R_{i}$ is the same for every ionic species ($R=1.5$ {\AA} in this work).
The ionic charges are point charges in the centers of hard spheres.
It was shown that this simple model can reproduce the non-monotonic concentration dependence of the activity coefficient of electrolytes as soon as we assume that the dielectric constant is concentration dependent \cite{vincze-jcp-inpress-2010}.
Since this work is a model calculation, we do not change the dielectric constant with the concentration, but we fix it at the value $\epsilon= 78.4$.

\begin{figure}[ht]
\vspace{5mm}
\centerline{\includegraphics*[width=7cm]{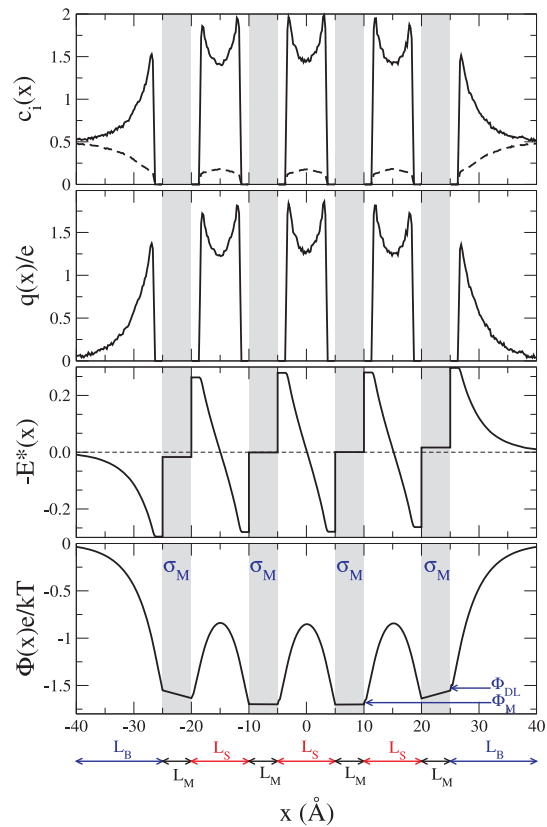}}
\caption{Geometry of the slit system and results for a 0.5 M 1:1 electrolyte in a slit system with parameters $N_{\mathrm{S}}=3$, $L_{\mathrm{S}}=10$ \AA, $L_{\mathrm{M}}=5$ \AA, and $\sigma_{\mathrm{M}}=-0.05$~Cm$^{-2}$. The concentration and charge profiles are measured in M, while the electric field profile is measured in \AA$^{-1}$. Here and in later figures, the membrane regions are indicated with shaded grey areas.}
\label{fig1}
\end{figure}
\vspace{5mm}

The $\alpha$th membrane is confined by two hard walls at $x_{\alpha}^{\mathrm{L}}$ and $x_{\alpha}^{\mathrm{R}}$, where each hard wall can carry a $\sigma_{\mathrm{M}} $ surface charge.
The interaction potential between such a charged hard wall and an ion is
\begin{equation}
v_{i}(|x|)=
\left\lbrace
\begin{array}{ll}
\infty & \quad \mathrm{for} \quad |x|<R_{i}\,, \\
 -\dfrac{z_{i}e\sigma_{\mathrm{M}}|x|}{2\epsilon_{0}\epsilon} & \quad \mathrm{for} \quad |x| \geqslant  R_{i}\,,
\end{array}
\right.
\label{eq:vi}
\end{equation}
where $|x|$ is the distance of the ion from the surface.

There are $N_{\mathrm{M}}$ membranes of width $L_{\mathrm{M}}$ (this is the distance of the two walls forming the membrane: $L_{\mathrm{M}}=x_{\alpha}^{\mathrm{R}}-x_{\alpha}^{\mathrm{L}}$ for every $\alpha$) in the slit system .
In this work, both membrane walls carry a $\sigma_{\mathrm{M}}$  surface charge.
The distance of two membranes, namely, the width of the slit is $L_{\mathrm{S}}$.
This distance is kept fixed during the simulation, namely, the slit system is rigid.
We should keep in mind, however, that the overlapping DLs between the like-charged macromolecules play a central role in the attractive force acting between them and contribute to the cohesion of such materials \cite{lyubartsev_prl_1998,allahyarov_jpcm_2005,pegado_jpcm_2008,zelko_jcp_2010}.

Ions are not allowed to enter the membranes, so the following Boltzmann factor is used to forbid the ions from the membranes:
\begin{equation}
 \exp \left(- \dfrac{u^{\mathrm{M}}_{\alpha,i}(x)}{kT} \right) =
\left\lbrace
\begin{array}{ll}
0 & \quad \mathrm{for} \quad x_{\alpha}^{\mathrm{L}}+R_{i}<x<x_{\alpha}^{\mathrm{R}}-R_{i}\,,\\
1 & \quad \mbox{otherwise},
\end{array}
\right.
\label{eq:um}
\end{equation}
where $x_{\alpha}^{\mathrm{L}}$ and $x_{\alpha}^{\mathrm{R}}$ are the coordinates of the left and right walls of the $\alpha$th membrane.
Obviously, the number of slits is $N_{\mathrm{S}}=N_{\mathrm{M}}-1$.

There are two bulk phases of widths $L_{\mathrm{B}}$ on the two sides of the slit system.
The simulation cell is closed by two hard walls on the left hand side of the left bulk region and on the right hand side of the right bulk region.
In this study, these walls are uncharged.
The geometry is illustrated in figure \ref{fig1}.

\section{Grand Canonical Monte Carlo simulations}
\label{sec:gcmc}

Grand Canonical Monte Carlo (GCMC) simulations have been performed for the system described above.
Periodic boundary conditions have been applied in the $y-z$ dimensions.
The effect of the ions outside the central simulation cell has been taken into account by the charged sheet method proposed by Torrie and Valleau \cite{torrie-jcp-1980} and developed further by Boda et.al. \cite{boda-jcp-109-7362-1998}.

In GCMC simulations of the DL, in addition to the usual particle displacement steps, we insert and delete neutral clusters of ions, e. g., $\nu_{+}$ cations and $\nu_{-}$ anions ($\nu_{+}$ and $\nu_{-}$ being the stoichiometric coefficients).
This way, we make sure that the simulation cell is charge neutral in every instant of the simulation.
The acceptance probability of these steps is
\begin{equation}
p^{\mathrm{in/out}}_{\pm} = \min  \left\lbrace 1, \dfrac{N_{+}!\,N_{-}!}{(N_{+}+\chi \nu_{+})! \,(N_{-}+\chi \nu_{-})!} V^{\chi\nu} \exp \left( \dfrac{-\Delta U +\chi \mu_{\pm}}{kT} \right)  \right\rbrace ,
\label{eq3:gcmc-ppm}
\end{equation}
where $N_{+}$ and $N_{-}$ are the numbers of cations and anions before insertion/deletion, $V$ is the volume into which we insert the centres of the ions, $\nu=\nu_{+}+\nu_{-}$, $\chi=1$ for insertion and $\chi=-1$ for deletion, while $\Delta U$ is the energy change of the insertion/deletion that contains the pair energies between ions, ions and charged walls, and ions and charged sheets (it becomes infinite in the case of overlap).
The quantity  $\mu_{\pm}=\nu_{+} \mu_{+}+\nu_{-}\mu_{-}$ is the chemical potential of the salt ($\mu_{+}$ and $\mu_{-}$ being the chemical potentials of the ions) determined by the Adaptive-GCMC (A-GCMC) method of Malasics et al. \cite{malasics-jcp-128-124102-2008,malasics-jcp-132-244103-2010}.

The dimensions of the simulation cell in the $y-z$ dimensions is in the range of 120--150~{\AA}.
%Note that although the average number of ions in a single slit might be small, their number is allowed to fluctuate in the GCMC simulation.
%Therefore, density fluctuations can be sampled adequately and the sampling problems
System-size checks indicated little sensitivity of the concentration profiles on the $y-z$ dimensions of the cell.
In a typical simulation, the sample contained several hundreds of millions ($10^{8}$) configurations.

\section{Solution of Poisson's equation}
\label{sec:poisson}

The main output quantities of the simulations are the density profiles of the various ionic species, $\rho_{i}(x)$, from which the ionic charge profile is obtained as
\begin{equation}
 q(x) = \sum_{i} z_{i}e \rho_{i}(x).
\label{eq:qdef}
\end{equation}
Poisson's equation
\begin{equation}
\dfrac{\rd^{2}\Phi (x)}{\rd x^{2}}= - \dfrac{1}{\epsilon_{0}\epsilon} q_{\mathrm{tot}}(x)
\label{poisson}
\end{equation}
is solved for the mean electrical potential $\Phi(x)$ using Neumann boundary conditions.
The total charge density $q_{\mathrm{tot}}(x)$ contains all charges in the system (including the $\sigma_{\mathrm{M}}$ membrane charges in addition to the ionic charge density $q(x)$):
\begin{equation}
 q_{\mathrm{tot}}(x) = q(x) + \sum_i \sigma_{i} \delta (x-x_i) ,
\end{equation}
where $ \sigma_{i}$ surface charges are placed in positions $x_{i}$.
After integration we obtain
\begin{equation}
\dfrac{\rd\Phi (x)}{\rd x} = - \dfrac{1}{\epsilon_{0}\epsilon} \int_{-\infty}^{x} q_{\mathrm{tot}}(x') \rd x' +C_{1}=
 -\dfrac{1}{\epsilon_{0}\epsilon} \int_{-\infty}^{x} q(x')\rd x' - \dfrac{1}{\epsilon_{0}\epsilon} \sum_{i (x_i<x)} \sigma_{i} + C_{1} \,,
\label{eq:ex}
\end{equation}
where the notation $i(x_{i}<x)$ under the sum means that we include only those sheets in the sum whose coordinates are smaller than $x$.
The numerical integration is performed using the rectangle rule.
The electric field is defined as
\begin{equation}
 E(x) = - \dfrac {\rd\Phi (x)}{\rd x}\, .
\end{equation}
Since our system is charge neutral:
\begin{equation}
\int_{-\infty}^{\infty} q(x)\rd x + \sum_{i}\sigma_{i}A = 0 ,
\end{equation}
the electric field outside the simulation domain is zero ($A=L^{2}$ is the area of the simulation cell in the $y-z$ plane).
The first boundary condition, therefore, is
\begin{equation}
 E(- \infty) = 0,
\end{equation}
from which the value $C_1=0$ is obtained for the first integration constant.
After one more integration the electrical potential is obtained
\begin{eqnarray}
 \Phi (x) &=&
- \dfrac{1}{\epsilon_{0}\epsilon}  \int_{-\infty}^{x} \left[ \int_{-\infty}^{x'} q_{\mathrm{tot}}(x'') \rd x''\right] \rd x'  + C_{1}x + C_{2} \nonumber \\
 &=& - \dfrac{1}{\epsilon_{0}\epsilon}  \int_{-\infty}^{x} \left[ \int_{-\infty}^{x'} q(x'') \rd x''\right] \rd x' + \left(- \dfrac{1}{\epsilon_{0}\epsilon} \sum_{i (x_i<x)} \sigma_{i} \right) x + C_{2}\,,
\label{eq:fix}
\end{eqnarray}
where we assumed that $C_{1}=0$.
Numerical integration is performed according to the trapezoidal rule.
Since the system is charge neutral, the Neumann boundary condition on the other side ($E(\infty)=0$) is automatically fulfilled.

The zero level of the potential is arbitrary. Therefore, we choose the integration constant $C_{2}$ so that the potential is zero in the left hand side bulk.
This is done by starting with $C_{2}=0$, averaging the potential over the left hand side bulk (where it is constant), and deducting this value from the potential profile.

Note that this method of solving the Poisson's equation is different from the traditional convolution integral
\begin{equation}
\Phi(x) = -\dfrac{1}{\epsilon_{0}\epsilon}\int_{x}^{L_{\infty}} (x'-x)q_{\mathrm{tot}}(x')\rd x',
\end{equation}
where the boundary condition is set in a bulk region ($L_{\infty}$).
This formula is used in theoretical studies of an isolated DL, where the boundary condition is set infinitely far from the electrode ($L_{\infty}\rightarrow \infty$, where the potential and the electric field are zero).
This equation can be used in simulation studies if there is a bulk region in the simulation cell and if the statistics is good enough.
If there is no bulk region, such as in the case of a lonely slit, the convolution integral is not practical.
Furthermore, equation \eqref{eq:fix} provides results for the potential profiles with much better accuracy than the convolution integral.
A formula that is equivalent to equation~\eqref{eq:fix} was introduced by Kiyohara and Asaka \cite{kiyohara07} without any reference to the fact that it corresponds to the Neumann boundary conditions.

\section{Results and discussion}
\label{sec:res}

In this work, we measure the distances in {\AA}, so particle densities are measured in {\AA}$^{-3}$.
In the figures, however, we plot concentration profiles that are related to the density profiles through $c_{i}(x)= 1660.58 \times \rho_{i}(x)$ (the unit of concentration is mol/dm$^{3}$ abbreviated as M).
The charge profile is also computed in terms of concentrations and is normalized by the elementary charge: $q(x)/\re=z_{+}c_{+}(x)+z_{-}c_{-}(x)$ (the unit is M).
The potential is plotted in a dimensionless form as $\re\Phi(x)/kT$.
The electric field is the derivative of this dimensionless potential, so its unit is {\AA}$^{-1}$.
We will denote it by $E^{*}(x)$.
\begin{figure}[!h]
\centerline{\includegraphics*[width=7cm]{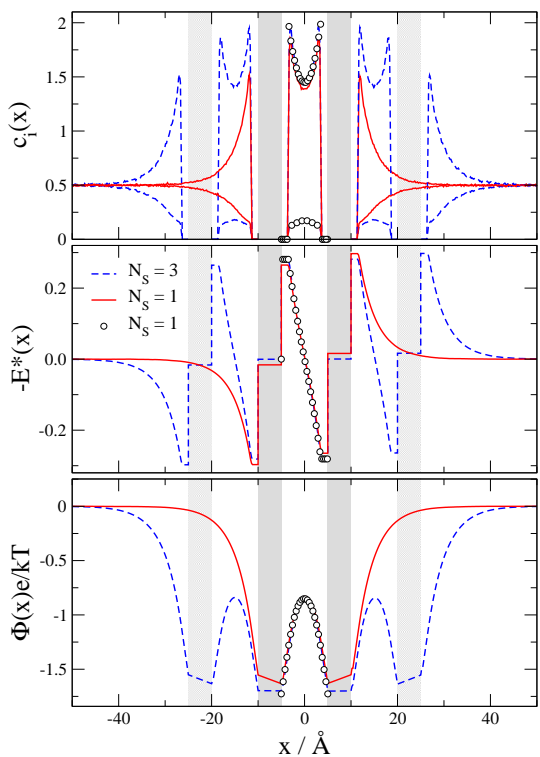}}
\caption{(Color online) The effect of the number of slits for a 0.5 M 1:1 electrolyte in slit systems with parameters $L_{\mathrm{S}}=10$ \AA, $L_{\mathrm{M}}=5$ \AA, and $\sigma_{\mathrm{M}}=-0.05$~Cm$^{-2}$. Results for $N_{\mathrm{S}}=1$ and 3 are shown. For the $N_{\mathrm{S}}=1$ case, results are also shown for the lonely slit, when the bulk regions are not included in the simulations (symbols). The outermost membrane region (shaded with lighter grey) is present only for the case $N_{\mathrm{S}}=3$, so they are not forbidden regions for the solid red line concentration profiles for $N_{\mathrm{S}}=1$ that seem to overlap with them. The role of the symbols is to distinguish the curve from other, overlapping curves; they indicate selected simulation points. This is also true for other figures, where symbols are used.}
\label{fig2}
\end{figure}

The temperature is $T=298.15$ K.
We used 1:1 and 2:1 RPM electrolytes with ions of radii $R=1.5$ {\AA}.
The size of the bulk phases was $L_{\mathrm{B}}=50$ {\AA} (simulations with $L_{\mathrm{B}}=80$ {\AA} gave the same results).
We show the results for various values of the number of slits ($N_{\mathrm{S}}$), width of slits ($L_{\mathrm{S}}$), width of membranes ($L_{\mathrm{M}}$), membrane charge ($\sigma_{\mathrm{M}}$), and electrolyte concentrations ($c$).

A typical result is seen in figure \ref{fig1}, where concentration, charge, electric field, and potential profiles are plotted for a 1:1 electrolyte adsorbed in a slit system with parameters indicated in the caption.
The concentration profiles of the cations (counter-ions) show peaks in the slits that are necessary to balance the membrane charges.
At these parameters, there is enough space in the slits to accept some coions too.
The concentration profiles of both ions approach the bulk values in the two bulk regions on the two sides of the slit system.
Near the external membranes double layers are formed.
These double layers are apparent in the charge profile.
They also balance the membrane charge but they extend into the bulk, because space is available to form diffuse layers.

\begin{wrapfigure}{i}{0.45\textwidth}
\centerline{\includegraphics*[width=0.43\textwidth]{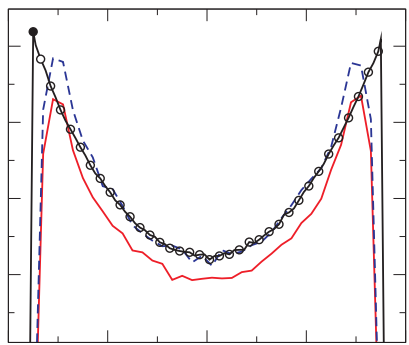}
\hspace{10mm}}
\vspace{-5mm}
\caption{The counter-ion concentration profiles of figure \ref{fig2} in the innermost slit.}
\label{fig3}
\end{wrapfigure}

These double layers are responsible for the potential drops between the bulk regions and the outmost membranes of the slit system ($\Phi_{\mathrm{DL}}$, see figure \ref{fig1}).
Let us define the membrane potential, $\Phi_{\mathrm{M}}$, as the potential difference between the inner membranes and the bulk.
The potential drop associated with the external DLs, $\Phi_{\mathrm{DL}}$, dominates the membrane potential, $\Phi_{\mathrm{M}}$.
The difference is due to the fact that there is less ionic charge in the outer slits than necessary to balance the membrane charge.
The missing charge in the outer DLs balances the membrane charges from outside.

The fact that the outer slits and the associated membrane charges do not cancel is proved by the electric field profile.
The electric field is zero in the inner membranes because inside the slit system the adsorbed ionic charge totally balances the membrane charge.
In the outer membranes, however, the electric field is non-zero indicating that there are more counter-ions in the outer DLs and less counter-ions in the outer slits.
(The non-zero electric field is reflected by the slope of the potential in the outer membranes.)
This polarization adds an additional term to $\Phi_{\mathrm{M}}$ denoted by $\Phi_{\mathrm{POL}}$.
Of course, the relation $\Phi_{\mathrm{DL}}+\Phi_{\mathrm{POL}}=\Phi_{\mathrm{M}}$ between these potential values holds.
In this case, $\Phi_{\mathrm{POL}}$ is much smaller than $\Phi_{\mathrm{DL}}$.

%\subsection{The effect of the number of slits}
%\label{sec:N_S}
\paragraph{The effect of the number of slits}
\label{sec:N_S}

Figure~\ref{fig2} shows the results for different values of $N_{\mathrm{S}}$ ($N_{\mathrm{S}}=1$ and 3), the lonely slit ($N_{\mathrm{S}}=1$, symbols), among them.
Only the outermost slit is polarized: the inner slits are not effected by the outer world.
The concentration and the potential profiles show a periodic behaviour in the inner slits.
This fact is even better demonstrated in later figures where we show the results for more slits.

An interesting result is that the behaviour of the lonely slit is quite similar to the behaviour of the innermost slit when bulk phases are present.
The shape of the potential profile is similar, but the reference point of the potential profile is not well defined.
At least, this cannot be related to the bulk region because the bulk region is not present in the simulation cell when the lonely slit is simulated.

\begin{figure}[ht]
\centerline{\includegraphics*[width=7cm]{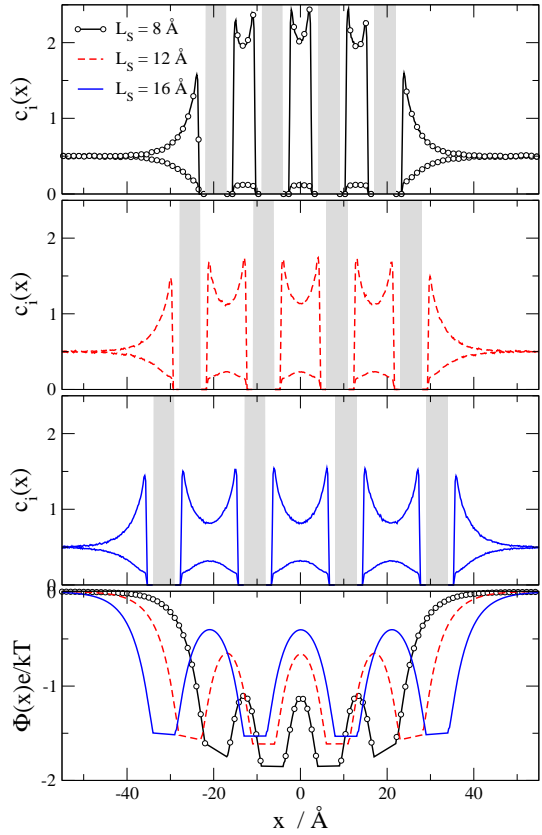}}
\caption{The effect of the width of slits for a 0.5 M 1:1 electrolyte in slit systems with parameters $N_{\mathrm{S}}=3$, $L_{\mathrm{M}}=5$ \AA, and $\sigma_{\mathrm{M}}=-0.05$~Cm$^{-2}$. Concentration profiles for values $L_{\mathrm{S}}=8$, 12, and 16 {\AA} are shown in the top panels, while the potential profiles are shown in the bottom panel.}
\label{fig4}
\end{figure}
\begin{figure}[!h]
\includegraphics[width=0.45\textwidth]{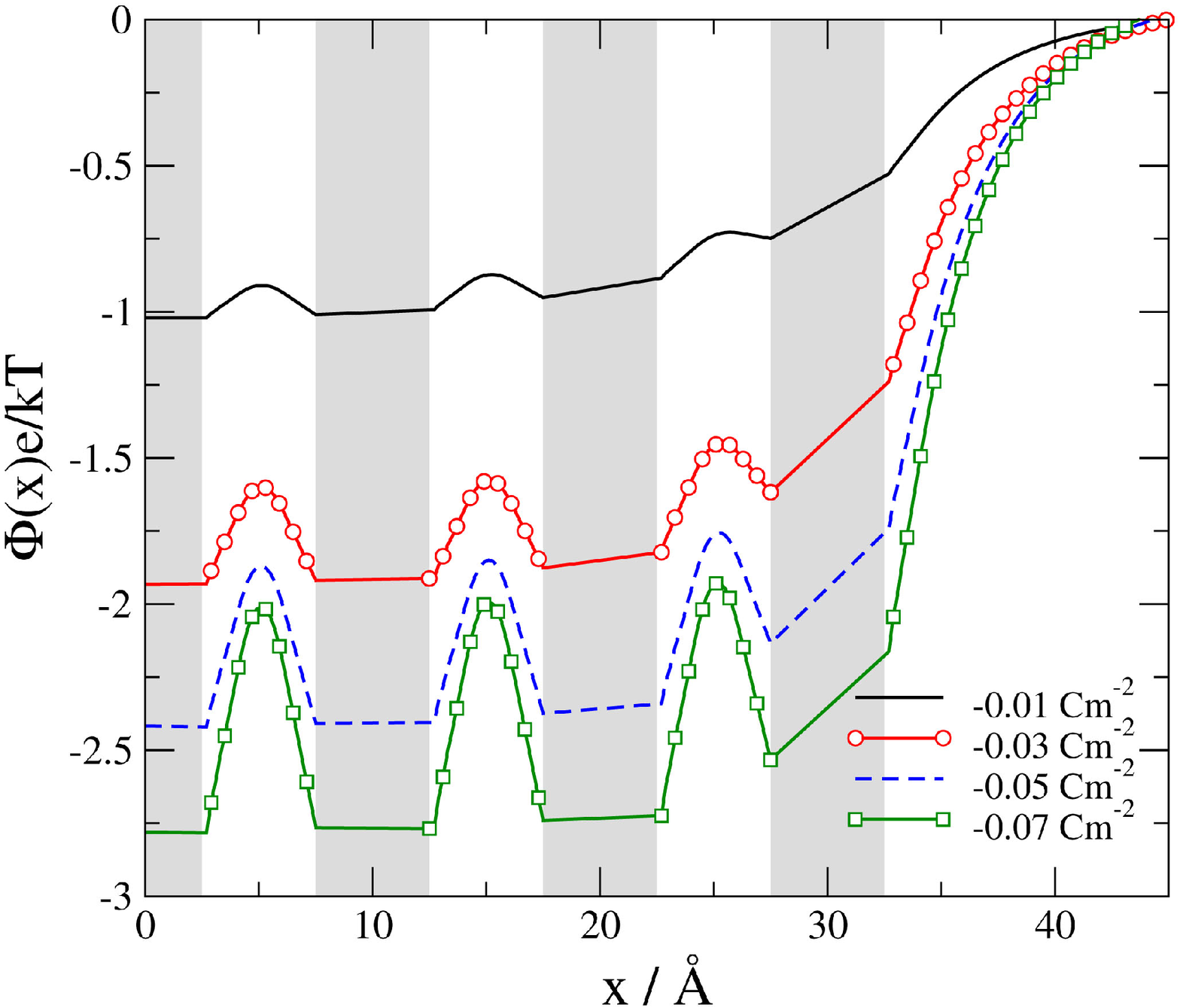}%
\hspace{1.5cm}%
\includegraphics[width=0.4\textwidth]{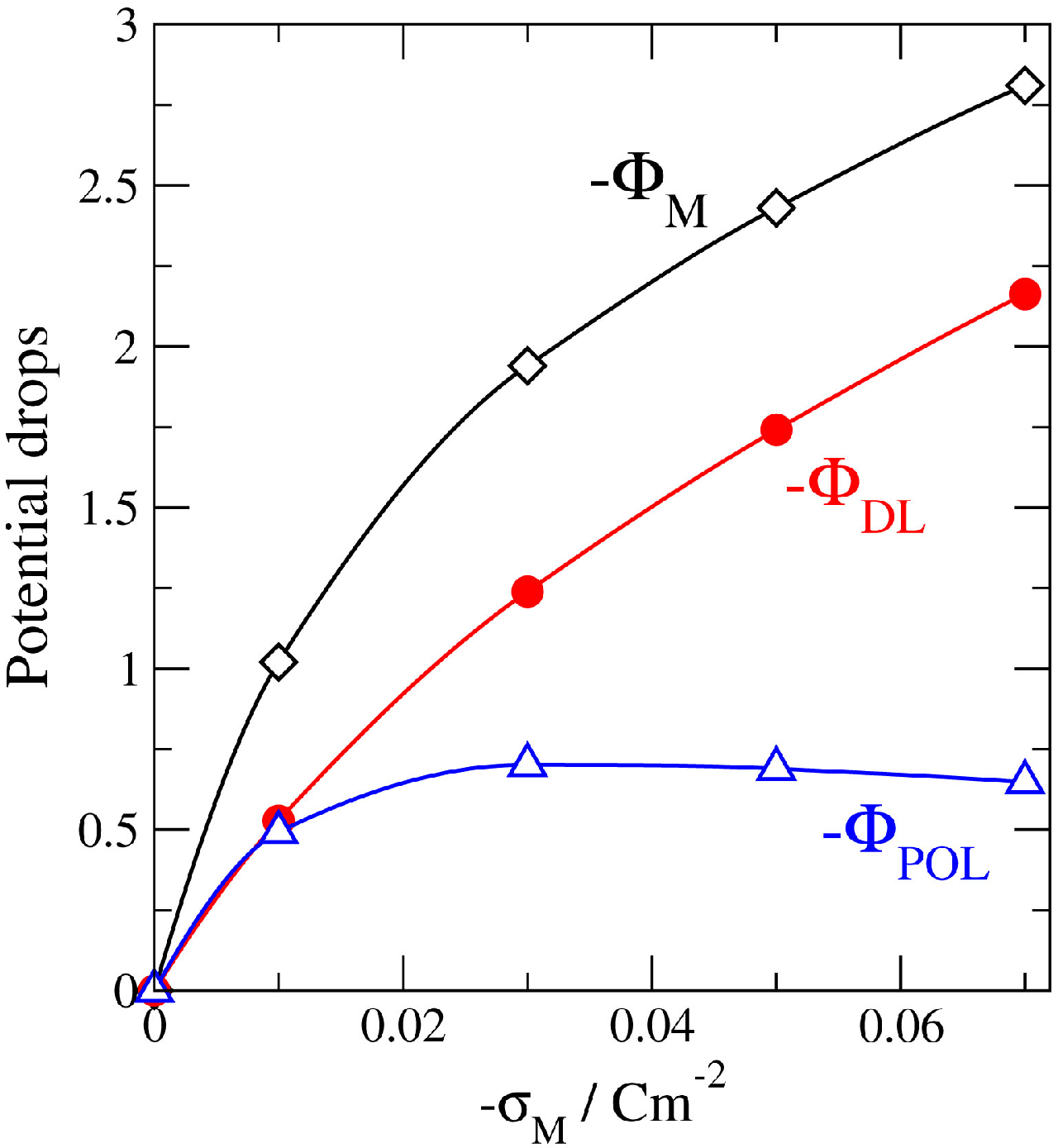}%
\\%
\parbox[t]{0.48\textwidth}{%
\caption{%
The effect of the membrane charge for a 0.5 M 1:1 electrolyte in a slit system with parameters $N_{\mathrm{S}}=6$, $L_{\mathrm{S}}=5$ \AA, and $L_{\mathrm{M}}=5$~\AA. Potential profiles for values $\sigma_{\mathrm{M}}=-0.01$, $-0.03$, $-0.05$, and $-0.07$~Cm$^{-2}$ are shown.}
\label{fig5}%
}%
\hfill%
\parbox[t]{0.48\textwidth}{%
\caption{The potential drops between the edge of the slit system and the bulk ($\Phi_{\mathrm{DL}}$) and between the inner membranes of the slit system and the bulk ($\Phi_{\mathrm{M}}$) based on the potential profiles of figure \ref{fig5}.}
\label{fig6}%
}%
\end{figure}
The shape of concentration profiles also look similar in the lonely slit and in the innermost slits, at least, at the scale of figure \ref{fig2}.
Figure~\ref{fig3} magnifies the peaks of the counter-ion profiles.
The ionic profiles in the lonely slit show a maximum at contact position: a typical hard sphere effect at the wall.
The ionic profiles in the slit system, on the other hand, show a smooth decrease at the wall.
This is an electrostatic effect: the dipole field of the polarized charges outside the slit system exerts a repulsive effect on the counter-ions.

This result shows that the lonely slit is a good approximation if we want to study the structure of the density profiles in the slit, but this is insufficient if we want to get information about the electrical work that is needed to bring an ion into the slit.
In that case, the slit should be simulated together with the bulk phases outside it.

%\subsection{The effect of the width of slits}
%\label{sec:L_S}
\paragraph{The effect of the width of slits}
\label{sec:L_S}

In the slits, the potential profile declines (in absolute value) to zero, but it does not reach zero because charge neutral bulk regions are not formed in the narrow slits.
If the slits were wide enough, independent DLs would form near the membranes and the potential would drop to zero.
Figure~\ref{fig4} shows this effect with an increasing slit width.

As $L_{\mathrm{S}}$ increases, the average concentration in the slits decreases and the cation and anion profiles in the middle of the slits become more and more similar.
The potential profile, therefore, gets closer to zero.
The potential drop in the DL, $\Phi_{\mathrm{DL}}$, is the same in all cases: the interior of the slit system does not effect  the outer DL.
The potential profile inside the slit system, on the other hand, sensitively depends on $L_{\mathrm{S}}$.
For large $L_{\mathrm{S}}$, the $\Phi_{\mathrm{POL}}$ potential drop vanishes, and the potential is the same in all membranes.
Polarization of the outer membranes ($\Phi_{\mathrm{POL}}$) appears for narrow slits.

%\subsection{The effect of the membrane charge}
%\label{sec:sigma}
\paragraph{The effect of the membrane charge}
\label{sec:sigma}

Increasing the charge of the membranes, $\sigma_{\mathrm{M}}$, increases the amount of ionic charge that is necessary to balance it.
This increases the charge difference between the slits and the membranes, which makes the potential fluctuation in the slit system larger (figure \ref{fig5}).
The amount of charge in the diffuse layers outside the slit systems also increases, which increases the potential drop across these DLs ($\Phi_{\mathrm{DL}}$).
The potential increases non-linearly with $\sigma_{\mathrm{M}}$ as seen in figure \ref{fig6}.
This phenomenon was already observed in the case of DLs \cite{boda-jcp-116-7170-2002}.

The DL and POL components of the potential drop are also shown in figure \ref{fig6}.
The relative weight of the DL component in $\Phi_{\mathrm{M}}$ is smaller at small membrane charges, while $\Phi_{\mathrm{POL}}$ is relatively small at large membrane charges.
As a matter of fact, $\Phi_{\mathrm{POL}}$ shows a non-monotonous behaviour and decreases with increasing $\sigma_{\mathrm{M}}$ at large membrane charges.

%\subsection{The effect of concentration}
%\label{sec:sigma}
\paragraph{The effect of concentration}
\label{sec:concentration}

\begin{figure}[ht]
\centerline{\includegraphics*[width=7cm]{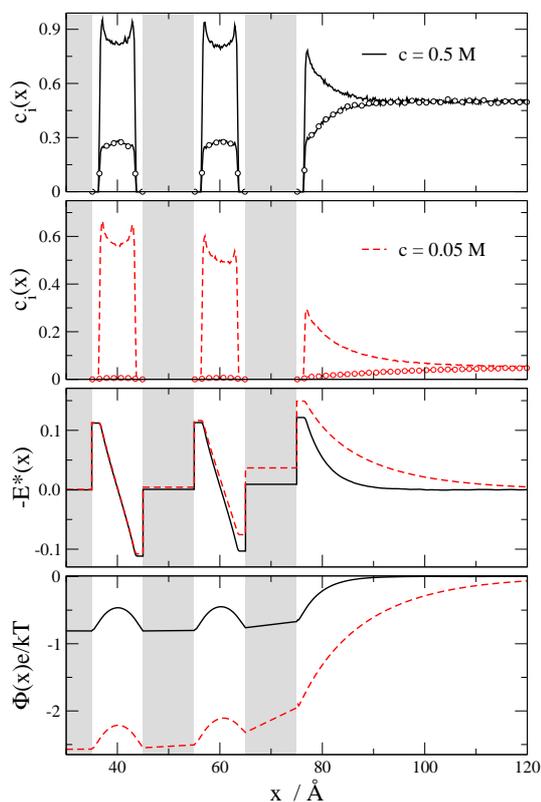}}
\caption{The effect of electrolyte concentration for 1:1 electrolytes in a slit system with parameters $N_{\mathrm{S}}=7$, $L_{\mathrm{S}}=10$ \AA, $L_{\mathrm{M}}=10$ \AA, and $\sigma_{\mathrm{M}}=-0.02$~Cm$^{-2}$. Concentration, electric field, and potential profiles for concentrations $c=0.05$ and 0.5 are shown. Only the outer slits at the right hand side of the slit system are shown. Anions are indicated with symbols.}
\label{fig7}
\end{figure}

The electrode potential of the DL layer for a given electrode charge (the capacitance of the DL) depends on the concentration of the electrolyte.
The electrical properties of the slit system also depend on the concentration.
 Figure~\ref{fig7} shows the difference between the results for 0.05 and 0.5 M electrolytes.
The net amount of ionic charge in a slit is fixed by the membrane charge, but this is achieved in different ways for the two concentrations.
The chemical potential $\mu_{\pm}$ that determines how many ions tend to be adsorbed in the slit are different at different concentrations.
At a higher concentration (0.5 M), there are more counter-ions in the slit and they drag some coions with themselves.
At a lower concentration (0.05 M), only counter-ions are in the slit.

The behaviour of the outer slits is also different in the two cases.
There are more missing charges in the outermost slit in the 0.05 M case.
This is also shown in the non-zero electrical field in the outermost membrane.
Therefore, the $\Phi_{\mathrm{POL}}$ potential term is larger at small concentrations.

The structure of the outer DLs is also different.
As in the case of a DL at an electrode, the $\Phi_{\mathrm{DL}}$ potential drop is larger at small concentrations.

%\subsection{The effect of divalent ions}
%\label{sec:sigma}
\paragraph{The effect of divalent ions}
\label{sec:divalent}

\begin{figure}[ht]
\centerline{\includegraphics*[width=7cm]{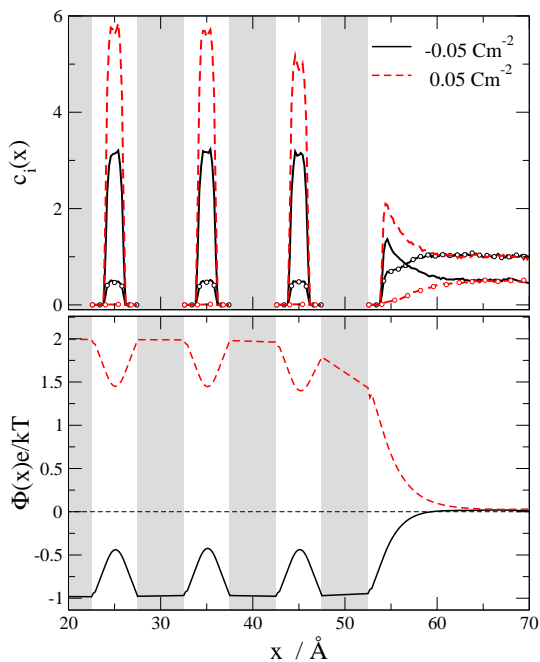}}
\caption{The effect of the sign of the membrane charge for a 2:1 electrolyte in a slit system with parameters $N_{\mathrm{S}}=10$, $L_{\mathrm{S}}=5$ \AA, and $L_{\mathrm{M}}=5$ \AA. Concentration and potential profiles for membrane charge $\sigma_{\mathrm{M}}=\pm0.05$~Cm$^{-2}$ are shown. Only the outer slits at the right hand side of the slit system are shown. Coions are indicated with symbols.}
\label{fig8}
\end{figure}

The overall situation for 2:1 electrolytes is very similar to that for 1:1 electrolytes.
The most notable difference occurs between membrane charges of the opposite sign but the same absolute value.
 Figure~\ref{fig8} shows results for a 2:1 electrolyte in a slit system with $\sigma_{\mathrm{M}}=\pm 0.05$~Cm$^{-2}$ membrane charges.

The divalent ions are more efficient in balancing the membrane charge than monovalent ions because they provide twice the charge occupying approximately the same space.
Polarization of the outer slit, therefore, is negligible at a negative membrane charge when the divalent cation is the counter-ion.
Since the divalent ions form a more compact diffuse layer, the potential drop $\Phi_{\mathrm{DL}}$ is smaller in this case.

At positive membrane charge, on the other hand, both $\Phi_{\mathrm{DL}}$ and $\Phi_{\mathrm{POL}}$ are larger (in absolute value) than in the case of negative membrane charge.

\section{Conclusions}
\label{sec:conclusions}

GCMC simulation results for PM electrolytes adsorbed in a slit system are reported.
We conclude that it is necessary to study the slit together with the bulk phase with which it is in equilibrium if we want to get  information on the potential difference between the slit and the bulk.

This potential difference has two main components.
The double layers outside the slit system produce potential drops analogous to the electrode potential in the case of a separate DL.
This component of the potential drop ($\Phi_{\mathrm{DL}}$) depends primarily on the properties of the bulk electrolyte.

If the ions have difficulty in entering the slits, the outermost slits have less charge than necessary to balance the membrane charge.
The membrane charge then is neutralized from outside, from the outer DLs.
This charge missing from the outer slits results in a polarization of these slits and in a potential difference.
This component of the potential drop ($\Phi_{\mathrm{POL}}$) depends on both the electrolyte properties and geometrical parameters of the slit system.
The $\Phi_{\mathrm{POL}}$ component is larger in absolute value (compared to the $\Phi_{\mathrm{DL}}$ component)  (1) if the width of the slits is smaller, (2) if the electrode charge is smaller in absolute value (but not too large), (3)  if the concentration is smaller, and (4) if the monovalent ion is the counter-ion (the membrane charge positive for 2:1 electrolytes).

We plan to study (1) competition between counter-ions of different charge and/or size (ionic selectivity -- this is inspired by our ion channel studies \cite{boda-prl-98-168102-2007}), (2) ions capable of being adsorbed into the membrane regulated by a Boltzmann-factor as in our previous membrane studies \cite{boda-jcp-111-9382-1999}, and (3) ion transport using our Local Equilibrium Monte Carlo method \cite{boda-jctc-submitted-2011}, with which a chemical potential gradient can be applied across the simulation cell.

\section*{Acknowledgements}

We acknowledge the support of the Hungarian National Research Fund (OTKA K75132).
Present publication was realized with the support of the project T\'AMOP--4.2.2/B--10/1--2010--0025.

\ukrainianpart

\title{Монте Карло симуляції електричних властивостей електролітів,
адсорбованих у заряджені щілиноподібні системи}
\author{Р. Ковач, М. Валіско, Д. Бода}
\address{
Факультет фізичної хімії, Університет Паннонії,
Веспрем, Угорщина
}

\makeukrtitle

\begin{abstract}
\tolerance=3000%
Використовуємо Монте Карло симуляції у великому канонічному ансамблі
для вивчення адсорбції примітивної моделі електролітів в шарувату
щілино-подібну систему. Остання містить низку заряджених мембран.
Мембрани є недоступними для іонів, але можуть адсорбуватись в
міжмембранні щілини. Ми концентруємо увагу на електричних властивостях
щілинної системи. Отримано концентрацію, заряд, електричне поле та
профіль електричного потенціалу. Показано, що потенціальна різниця між
щілинною та об'ємною фазами спричинена, в основному, виникненням на
границях щілини подвійних шарів, але вклад вносить також і поляризація
зовнішніх щілин. Ми демонструємо, що електрична робота, яку потрібно
здійснити з метою внесення іона в щілинну систему, може бути отримана
лише тоді, коли симуляція щілини виконується разом із об'ємними фазами
в одній симуляційній комірці.
\keywords Монте Карло, примітивна модель електролітів, щілини

\end{abstract}

\end{document}